# LiYbSe$_2$: Frustrated Magnetism in a New Pyrochlore Lattice

Ranuri S. Dissanayaka Mudiyanselage[1], Haozhe Wang[1], Olivia Vilella[2], Martin Mourigal[2], Gabriel Kotliar[3], Weiwei Xie[1*]

[1]*Department of Chemistry and Chemical Biology, Rutgers University, Piscataway, New Jersey, 08854, USA*

[2]*School of Physics, Georgia Institute of Technology, Atlanta, Georgia 30332, USA*

[3]*Department of Physics and Astronomy, Rutgers University, Piscataway, New Jersey 08854, USA*

*Abstract*

Three-dimensionally (3D) frustrated magnets generally exist in the magnetic diamond and pyrochlore lattices, in which quantum fluctuations suppress magnetic orders and generate highly entangled ground states (GS). LiYbSe$_2$ in a previously unreported pyrochlore lattice was discovered from LiCl flux growth. Distinct from the quantum spin liquid (QSL) candidate NaYbSe$_2$ hosting a perfect triangular lattice of Yb$^{3+}$, LiYbSe$_2$ crystallizes in the cubic pyrochlore structure with space group F*d*-3*m* (No. 227). The Yb$^{3+}$ ions in LiYbSe$_2$ are arranged on a network of corner-sharing tetrahedra, which is particularly susceptible to geometrical frustration. According to our temperature-dependent magnetic susceptibility measurements, the dominant antiferromagnetic interaction in LiYbSe$_2$ is expected to appear around 8 K. However, no long-range magnetic order is detected in thermomagnetic measurements above 70 mK. Specific heat measurements also show magnetic correlations shifting with applied magnetic field with a degree of missing entropy that may be related to the slight mixture of Yb$^{3+}$ on the Li site. Such magnetic frustration of Yb$^{3+}$ is rare in pyrochlore structures. Thus, LiYbSe$_2$ shows promises in intrinsically realizing disordered quantum states like QSL in pyrochlore structures.

Frustrated magnetic systems, such as QSL and spin ice, are usually associated with highly entangled quantum states and spin excitation continuum, which have attracted much attention in materials science.[1–4] Spin excitation continuum can be observed in the geometrically frustrated spin-1/2 (S= ½) systems, including, two-dimensional (2D) Kagomé, triangular, 3D distorted Kagomé, and 3D pyrochlore lattices.[4–6] Rare-earth-based frustrated magnetic systems, such as YbMgGaO$_4$, a Yb$^{3+}$ with an effective S= ½ triangular lattice compound, have recently been proposed as a promising QSL candidate.[7] However, the intrinsic structural disorder with a random distribution of Mg$^{2+}$ and Ga$^{3+}$ in YbMgGaO$_4$ may generate a spin-liquid-like state at low temperatures. Thus, NaYbCh$_2$ (Ch=O, S, and Se), which contain perfect Yb$^{3+}$ triangular lattice, have been widely explored as an S= ½ system without inherent atomic disorders in crystals.[8–13]

Recent studies on single crystalline NaYbSe$_2$ have suggested a QSL GS with a spinon Fermi surface.[10] Considering that Na$^+$ ionic size is larger than Yb$^{3+}$ ($R_{Na^+}$ ~1.16 Å; $R_{Yb^{3+}}$ ~0.99 Å), we attempted to use Li$^+$ to replace Na$^+$ ($R_{Li^+}$ ~0.90 Å)[14] to reduce the chemical disorder in NaYbSe$_2$. Interestingly, instead of forming the perfect triangle Yb$^{3+}$ lattice in rhombohedral delafossite structure, LiYbSe$_2$ crystallizes in a cubic pyrochlore structure with the Yb$^{3+}$ tetrahedral lattice. Generally, the crystal structure of pyrochlore structure can be formulated as A$_2$B$_2$O$_6$ and A$_2$B$_2$O$_7$ where the A and B species are usually rare-earth or transition-metals occupying 16$d$ and 16$c$ sites while O atoms locate on 48$f$ site (and 8$b$ site). In LiYbSe$_2$, Li and Yb still occupy the 16$d$ and 16$c$ sites. However, Se atoms are on 32$e$ site due to their larger atomic size than O atom. Pyrochlores systems comprising rare-earth ions are a known avenue to explore anisotropic frustrated magnetism.[15,16]

Herein, we present the discovery of a new member of the alkali-metal rare-earth dichalcogenide family and a thorough structural characterization and investigation of the thermos-magnetic properties of LiYbSe$_2$, in which Yb$^{3+}$ adopts an effective S= ½ frustrated pyrochlore lattice.

**New Pyrochlore Structure of LiYbSe$_2$:** The crystallographic data analysis by single-crystal X-ray diffraction reveals that the compound, LiYbSe$_2$ crystallizes in the space group *Fd-3m*. As shown in **Figure 1a and 1b,** all octahedral voids are occupied by Yb$^{3+}$ and Li$^+$ which are arranged in alternate layers of Li$^+$ and Yb$^{3+}$ along [111] direction. The new LiYbSe$_2$ pyrochlore structure type can be derived from the rare-earth (RE) transition metal chalcogenides, (RE$^{3+}$)$_2$M$^{2+}$(Ch$^{2-}$)$_4$ that adopts 3D cubic spinel structure where the rare-earth atom is in the octahedral geometry and the transition-metal in tetragonal geometry (**Figure S1**). By replacing M$^{2+}$ with a much smaller Li$^+$, transformation to octahedral geometry in both Yb and Li atoms is observed to charge balance and accommodate close packing. In LiYbSe$_2$, Li atoms prefer the 16*c* site which was occupied by rare-earth metal in the cubic spinel structure type while Yb atom in LiYbSe$_2$ resides on the 16*d* site. Moreover, the Yb atoms on the 16*d* site are arranged into corner-sharing tetrahedra extending to a pyrochlore sublattice (**Figure 1c**) which can be also viewed as a 3D expansion of the Kagome lattice. Yb atoms in the center of the triangle are connecting the alternating layers of Kagome extending to 3D as shown in **Figure 1d**. Alkai-metal rare-earth dichalcogenides are a large class of compounds that crystallizes in different structure types depending on the different sizes of the cations and anions. Several crystal space groups are reported in the system, such as LiLaO$_2$ in *P*2$_1$/*c* (No. 14), LiEuO$_2$ in *Pnma* (No. 62), LiYbO$_2$ in *I*4$_1$/*amd* (No. 141), NaErO$_2$ in *C*2/*c* (No. 15), NaYbSe$_2$ in *R-3m* (No. 166), CsTbSe$_2$ in *P*6$_3$/*mmc* (No. 194) and NaLaS$_2$ in *Fm-3m* (No. 225).[17–23] The trends of the crystal structure changes relative to the ratio rare-earth ion radius to alkali plus chalcogenide radii (RE$^{3+}$/(Ch$^{2-}$ + A$^+$)), has been summarized for the A(RE)O$_2$, A(RE)S$_2$, A(RE)Se$_2$ and A(RE)Te$_2$ series.[17]

Accordingly, even though A(RE)O$_2$ exhibits to crystallize in a wide range of structure types as mentioned above, A(RE)S$_2$, A(RE)Se$_2$ and A(RE)Te$_2$ mostly crystallize in *R-3m* and *P*6$_3$/*mmc* and occasionally in *C*2/*c*.[17] In fact, within this broad family of A(RE)Ch$_2$, LiYbSe$_2$ is the first candidate to crystalize in the cubic *Fd-3m* with $J_{eff}$ =1/2. The only other compound reported in the family with the same space group is the NaPrTe$_2$.[24] The Rietveld refinement of powder X-ray diffraction (PXRD) shown in **Figure 1e** confirms that synthetic attempts have yielded pure LiYbSe$_2$.

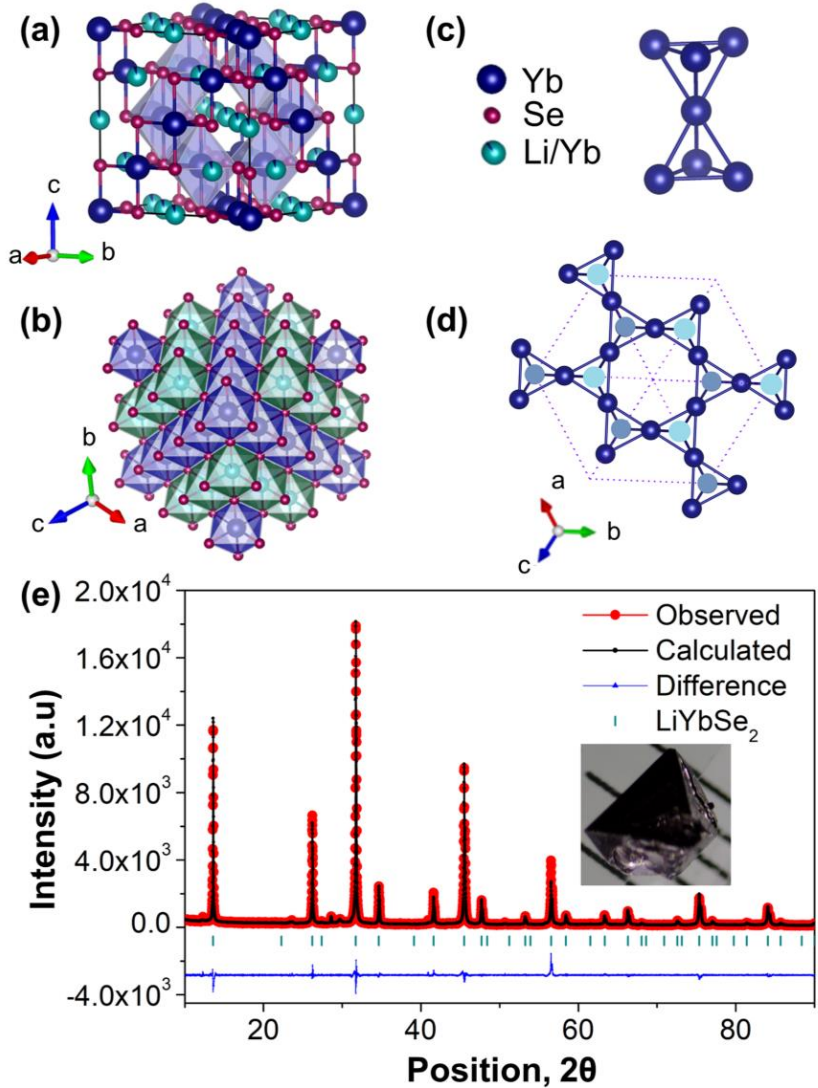

**Figure 1.** (*a*) The unit cell of LiYbSe$_2$. (*b*) Li$^+$ and Yb$^{3+}$ octahedra along [111] direction (*c*) The pyrochlore sublattice of Yb$^{3+}$ (*d*) Bulk Kagome lattice (*e*) PXRD of LiYbSe$_2$ and diamond shaped crystal picture of LiYbSe$_2$.

**Magnetic Properties of LiYbSe$_2$:** As represented in **Figure 2***a*, the field-dependent magnetization measurements show a non-linear behavior at 2 K without a complete magnetic saturation below the applied field of $\mu_o H$=9 T. Therefore, magnetization data measured above 9 T is required to determine Van-Vleck contribution to magnetism. Subsequently, the observed maximum magnetization, M$_s$ = 1.3 μ$_B$/Yb$^{3+}$ at 9 T is lower than the powder average *g*-factor ($g_{avgCW}$/2 ~ 1.9) derived from Curie-Weiss analysis of magnetic susceptibility.

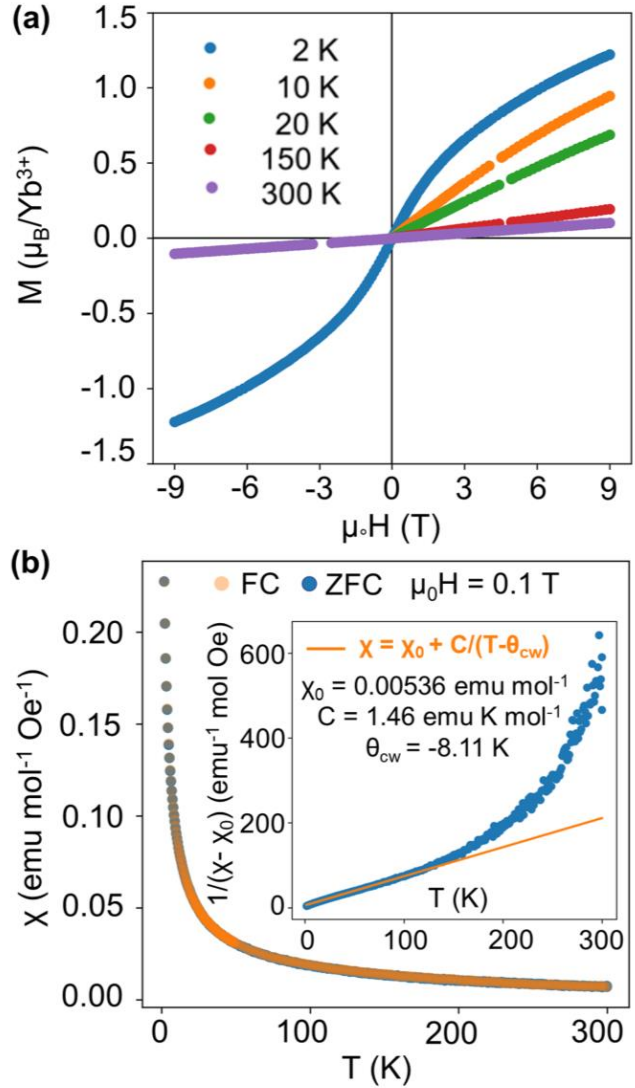

**Figure 2(*a*)** Magnetization isotherms measured at various temperatures. (***b***) Magnetic susceptibility at 1000 Oe using both zero field cooled (ZFC) and field cooled (FC) method. ***Inset:*** The Curie-Weiss fitting of inverse magnetic susceptibility at temperature range 20 – 100 K.

The temperature dependence of the magnetization in both ZFC and FC modes are presented in **Figure 2b** with no significant difference. The magnetic susceptibility ($\chi$) shows no long-range magnetic order above 1.8 K and no hysteresis was observed at 2 K. These observations are incompatible with long-range or short-range spin glass or spin ice states.[25,26] In the low-temperature area, the Kramers doublet GS is primarily occupied and the Van-Vleck contribution to the susceptibility data is negligible. The $1/(\chi-\chi_0)$ data from 20 K to 100 K (***Inset* of Figure 2b**) is fitted by modified Curie-Weiss law. The results yield a local moment of 3.42 $\mu_B$ with a Curie-Weiss temperature, $\theta_{CW}$ = -8.11 K. The negative $\theta_{CW}$ obtained for the low-temperature region reveals dominant antiferromagnetic interactions in the material. Also, the Curie-Weiss temperature is substantially enhanced compared with $\theta_{CW}$ = -4 K in YbMgGaO$_4$, which is consistent with the unsaturated polarized Yb moments and probably due to the enhanced exchange in the system.[27] Furthermore, results implies a powder-averaged *g* factor $g_{avgCW}$ = 3.95 assuming $J_{eff}$ = ½ for Yb ions. Continuously, we explored the low-temperature magnetic behavior of LiYbSe$_2$. The Curie-Weiss fitting of χ(T) between 1.8 K and 10 K (**Figure S3 and Table S4)**, produces a linear fit with an average moment 2.94 $\mu_B$ and $\theta_{CW}$ = -3.14 K. The coordination environment and consequent crystal electric field splitting plays a key role in the temperature dependent magnetization of the rare-earth elements. The resulting experimental $\mu_{eff}$ at low temperature for octahedral coordinated Yb$^{3+}$ materials tend to show lower values for example, as in KbaYbB$_2$O$_6$.[28,29] Above 100 K, the nonlinearity of the magnetic susceptibility arises, attributed to Van-Vleck contributions that originates from the crystalline electron field splitting of the $J = 7/2$ Yb manifold. The high temperature data fitting **(Figure S3)** between 200 K and 300 K yields $\mu_{eff}$ = 4.18 $\mu_B$, which is comparable to the $\mu_{eff}$ value of the theoretical prediction for the $^2F_{7/2}$ multiplet Yb$^{3+}$ ion (4.54 $\mu_B$).

To further evaluate the low-temperature thermodynamics of LiYbSe$_2$, specific heat ($C_p$) measurements were performed down to 70 mK. The $C_p$ consists of several contributions named nuclear, phonon, and magnetic which can be expressed as $C_p(T) = C_{nuc}(T) + C_{phonon}(T) + C_{mag}(T)$. The $C_p$ data was analyzed over three temperature regimes, ultra-low (ULT) (0.07 K < $T$ < 1.0 K), low (LT) (1.0 K < $T$ < 2.0 K) and high temperature (HT) (2.0 K < $T$ < 20 K) to understand the contribution from each component.

As indicated in **Figure 3***a*, the upturn observed in ULT range under the applied fields originates from the nuclear contribution of Yb nuclei, mainly from $^{171}$Yb and $^{173}$Yb. This can be described by $C_{nuc} = \alpha T^{-2}$, where $\alpha$ is a field dependent coefficient. The effect becomes sizable only below ~0.3 K for Yb nuclie.[30] In ULT range, $C_{phonon}$ contribution is negligible and $C_{mag}$ contribution is considered in the form of power law $C_{mag} = bT^p$ to account for gapless excitations. Details of fitted parameters for $C_p$ at different fields are summarized in the **Table S5**, **Figure S***4,* **and S5.** The fitted power exponent, *p* increases up to ~2.5 under 12 T, suggesting that LiYbSe$_2$ is a possible strongly correlated QSL candidate similar to YbMgGaO$_4$.[27] However, the zero-field data appeared to be more complicated, as data did not follow the power law well. Therefore, to determine $C_{nuc}$ we fitted the data with two power laws, $bT^p + dT^q$ where *p*=2.29 and *q*=0.683. While we tentatively fitted the data with two power law, there may be also other functions to fit the experimental data. The resultant model fit is represented in **Figure 3b**, which yields the coefficient α~3.25×10$^{-4}$ J-K/mol. The value is comparable with frustrated triangular materials NaYbSe$_2$ and NaYbO$_2$.[12,13] Further theoretical investigations will indeed require to fully understand this unusual zero-field $C_{mag}$ behavior at ULT.

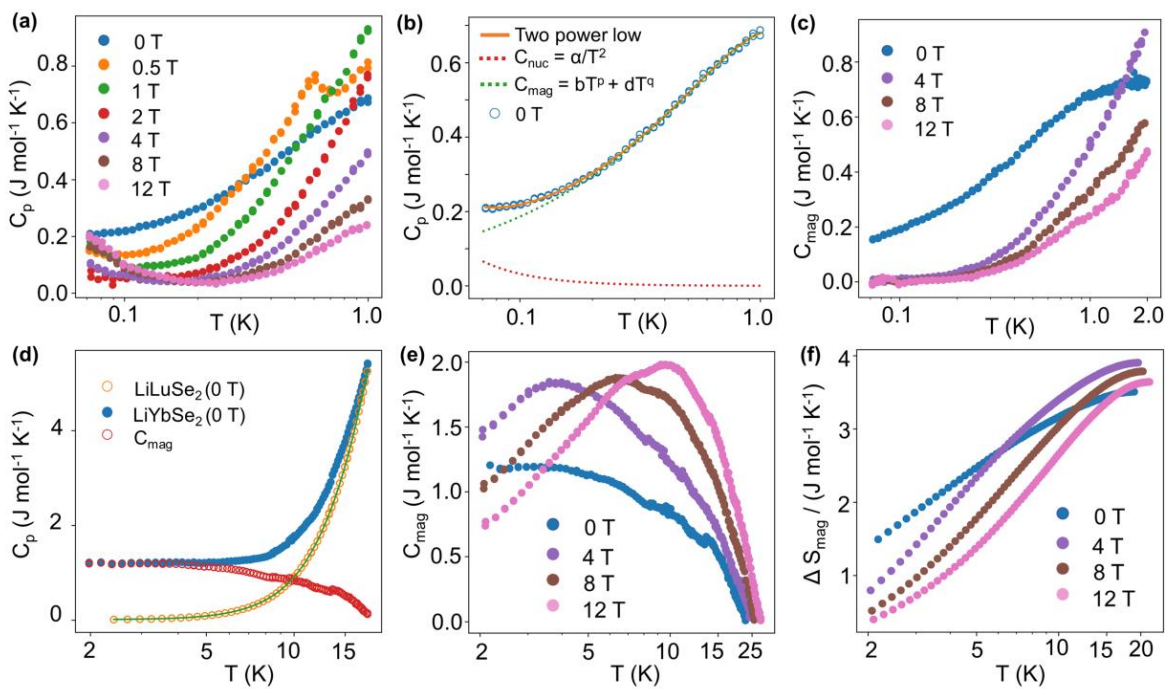

**Figure 3.** (*a*) $C_p$ measured at multiple fields for LiYbSe$_2$ at ULT. (*b*) Zero-field $C_p$ analysis at ULT regime (*c*) Calculated $C_{mag}$ for LiYbSe$_2$ at ULT+LT range (*d*) $C_{mag}$(0 T) analysis at HT with non-magnetic reference compound LiLuSe$_2$ (*e*) $C_{mag}$ at HT under different fields (*f*) Cumulative magnetic entropy versus temperature up to 20 K.

In LT range, the nuclear contribution decays enough, and the phonon contribution remains too small to be considered. Then the total specific heat is equivalent to $C_{mag}$ and the measured $C_p$ data is directly used for magnetic entropy analysis and results are given in **Figure 3c**. Finally, in the HT range, lattice contribution is non-negligible and $C_p(T) = C_{phonon}(T) + C_{mag}(T)$. **Figure 3d** shows the zero-field $C_p$ of LiYbSe$_2$ together with the non-magnetic reference compound LiLuSe$_2$. $C_{mag}$ has been precisely evaluated by subtracting the $C_{phonon}$ derived from $C_p$ of isostructural LiLuSe$_2$. As indicated in **Figure 3c and 3e**, a broad peak in $C_{mag}$ appeared as the temperature decreases, while the peak position is field dependent and shifts to higher temperatures with increasing magnetic fields, likely indicative of the onset of short-range correlations in the system. This behavior implies a crossover into a QSL which also observed in QSL candidate YbMgGaO$_4$.[27]

Furthermore, consistent with magnetic susceptibility data discussed previously, sharp anomaly representative of the onset of long-range order in the system was not observed down to 70 mK. All these evidences suggests a possible ideal QSL GS for LiYbSe$_2$ compared to other frustrated systems LiYbS$_2$, NaYbSe$_2$, and NaYbO$_2$ due to its higher frustration parameter ($f = \theta_{CW}/T_C$; 8/<0.07).[11,13,31] The calculated magnetic entropy ($\Delta S_{mag}$) and the field dependance with varying fields of 4 T, 8 T and 12 T is given in **Figure 3f** and observed that $\Delta S_{mag}$ does not reach Rln2 limit which is expected for an effective spin, S$_{eff}$ = ½ up to T~20 K. This indicates that there is residual entropy at 70 mK while whether this would be enough for a further magnetic transition remain unknown. However, no transition was observed down to the measured temperature 70 mK. Additionally, LiYbSe$_2$ is an excellent insulator with a resistance greater than 25 MΩ at room temperature. These results denote that LiYbSe$_2$ is an ideal QSL candidate. Interestingly, despite its experimental insulating state, fundamental electronic structure (**Figure S9**) calculation suggests a semi-metallic state for LiYbSe$_2$. This indicates the complexity of its electronic behavior in the new pyrochlore structure that requires future theoretical impact to fully understand its exotic properties.

In summary, new LiYbSe$_2$ with a geometric frustrated Yb$^{3+}$ pyrochlore sublattice was discovered. The thermomagnetic results show no spin freezing down to 70 mK despite the significant antiferromagnetic exchange energy ($\theta_{CW}$ = -8 K), suggesting a QSL state. The higher frustration index and field dependent positive shift of broad peak in $C_{mag}$ further supports this behavior. LiYbSe$_2$ is not only interesting as a QSL candidate, but its unusual zero-field $C_{mag}$ behavior at ULT and electronic structure provide an avenue to expand the frontiers of our knowledge on quantum materials into new areas.

## Acknowledgement

The work at Rutgers was supported by the U.S. Department of Energy (DOE), Office of Science, Basic Energy Sciences under award DE-SC0022156. The work at Georgia Tech was supported by NSF-DMR-1750186.

# Supporting Information

**Experiment and Calculation Details**

**Single Crystal Synthesis:** The LiYbSe$_2$ was synthesized using the high-temperature flux growth method. The elemental constituents Yb powder (99.9%, Alfa Aesar), Se shots (99.999%, BTC), and the flux LiCl (98+%, Alfa Aesar) were measured in the molar ratio of 1:2.2:20. The chemicals were handled and packed in the argon-filled glovebox to avoid oxidation of Yb and the moisture absorption of LiCl. The weighed mixture of Yb and Se was sandwiched in between the LiCl layers inside the alumina crucible. The crucible was placed inside the quartz tube and covered with quartz wool and sealed into an evacuated tube. Then the evacuated tube was heated to 400 °C annealed for 3 hours slowly ramped to 850 °C and annealed for one week before cooling down to room temperature at a rate of 15 °C/hr. The sample was washed with distilled water to clean the excess flux and air-dried before using it for further measurements. The property measurements were done on the polycrystalline sample. The non-magnetic reference compound, LiLuSe$_2$ was successfully synthesized using the same procedure and used for heat capacity measurements.

**Single Crystal X-ray Diffraction (SCXRD):** Several single crystals were picked up and conducted the SCXRD experiments in the Bruker D8 Quest Eco single-crystal X-ray diffractometer equipped with Photon II detector and Mo radiation (($\lambda_{K\alpha}$ = 0.71073 Å). The crystal was mounted on the tip of the Kapton loop and measured with an exposure time of 10 s and a scanning 2θ width of 0.5° at room temperature. The data acquisition, extraction of intensity, and correction for Lorentz and polarization effects were performed in Bruker Apex III software. The structural solutions were obtained with the SHELXTL package using direct methods and refined by full-matrix least-squares on F$^2$.[1,2] The crystallographic data analysis reveals that the compound, LiYbSe$_2$ crystallizes in the space group *Fd-3m* (No. 227) with the lattice parameter 11.2421(11) Å. Yb and Li sites are surrounded by six Se atoms with bond lengths 2.807(2) Å and 2.815(2) Å respectively. The detailed crystallographic data including atomic coordinates, site occupancies, and equivalent isotropic thermal displacement parameters are reported in **Table S1, S2 and S3.** The distribution of Li and Yb cations was examined by refining the occupancy parameter and revealed that the Li site (16*c*) requires a contribution from a heavy atom while the Yb site is fully occupied. Therefore, a mixture of Li/Yb was tested on Li (16*c*) site. The results indicate that Li site is mixed with ~5.8(4) % of Yb atoms and a satisfying agreement factor of R$_1$ = 5.9% was obtained. Furthermore, the Se site (32*e*) does not show vacancies or mixed occupancies. Accordingly, the detailed composition of the compound is [Li$_{0.94(1)}$Yb$_{0.06(1)}$]YbSe$_2$. The classic pyrochlore structures $A_2B_2X_7$ where both A and B form separate pyrochlore sublattices also show similar mixed occupancy effects as in the case of LiYbSe$_2$. However, the degree of mixed occupancy in LiYbSe$_2$ (~6%) is lower compared to some pyrochlore materials which can even go up to 60%.[3,4]

**Table S1.** Single crystal structure refinement for LiYbSe$_2$ at 300 (2) K. (Standard deviation is indicated by the values in parentheses)

| Refined Formula | Li$_{0.94(1)}$Yb$_{1.06(1)}$Se$_2$ |
|---|---|
| F.W. (g/mol) | 347.76 |
| Space group; Z | *Fd-3m*; 16 |
| *a*(Å) | 11.2421 (11) |
| V (Å$^3$) | 1420.8 (4) |
| θ range (°) | 3.139-34.924 |
| No. reflections; $R_{int}$ | 2533; 0.0723 |
| No. independent reflections | 182 |
| No. parameters | 9 |
| $R_1$: $\omega R_2$ ($I>2(I)$) | 0.0401;0.1095 |
| Goodness of fit | 1.348 |
| Diffraction peak and hole (e$^-$/ Å$^3$) | 1.675; -1.534 |

**Table S2.** Atomic coordinates and equivalent isotropic displacement parameters of LiYbSe$_2$ at 300 (2) K. (U$_{eq}$ is defined as one-third of the trace of the orthogonalized U$_{ij}$ tensor (Å$^2$)).

| Atom | Wyckoff. | Occ. | *x* | *y* | *z* | U$_{eq}$ |
|---|---|---|---|---|---|---|
| Yb1 | 16*d* | 1 | ½ | ½ | ½ | 0.0079(3) |
| Se2 | 32*e* | 1 | 0.2504(1) | 0.2504 (1) | 0.2504(1) | 0.0106(4) |
| Li3 | 16*c* | 0.941(9) | 0 | 0 | 0 | 0.046(11) |
| Yb4 | 16*c* | 0.059(9) | 0 | 0 | 0 | 0.046(11) |

**Table S3.** Anisotropic thermal displacement parameters of LiYbSe$_2$.

| Atom | U11 | U22 | U33 | U23 | U13 | U12 |
|---|---|---|---|---|---|---|
| Yb1 | 0.0079(3) | 0.0079(3) | 0.0079(3) | -0.0015(2) | -0.0015(2) | -0.0015(2) |
| Se2 | 0.0106(4) | 0.0106(4) | 0.0106(4) | 0.0053(5) | 0.0053(5) | 0.0053(5) |
| Li3 | 0.046(11) | 0.046(11) | 0.046(11) | 0.036(10) | 0.036(10) | 0.036(10) |
| Yb4 | 0.046(11) | 0.046(11) | 0.046(11) | 0.036(10) | 0.036(10) | 0.036(10) |

**Figure S1.** (*a*) The LiYbSe$_2$ pyrochlore structure type with Yb$^{3+}$ and Li$^+$ occupying the octahedral voids (*b*) MnYb$_2$Se$_4$ representing rare-earth (RE) transition metal chalcogenides, (RE$^{3+}$)$_2$M$^{2+}$(Ch$^{2-}$)$_4$ with the three-dimensional cubic spinel structure where the rare earth atom is in the octahedral geometry and the transition metal in tetragonal geometry.

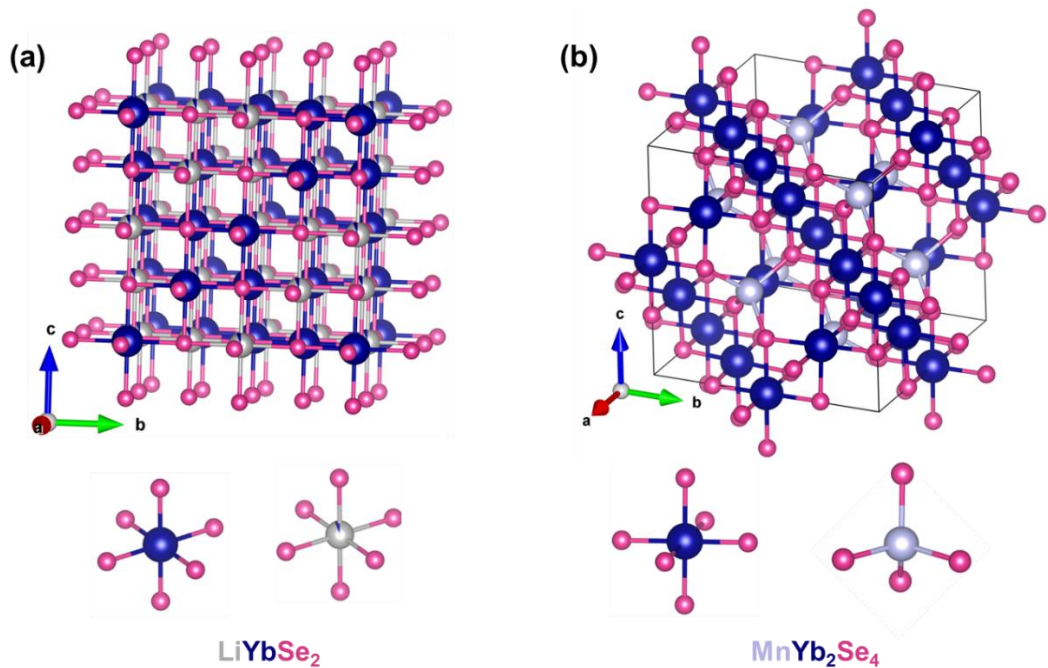

**Powder X-ray Diffraction (PXRD):** The phase determination of LiYbSe$_2$ and LiLuSe$_2$ samples were performed by the Bruker D2 Phaser XE-T edition benchtop X-ray Powder Diffractometer equipped with Cu K$_\alpha$, λ=1.5405 Å. Data were collected over a range of Bragg angle, 2θ, 5 to 90° at a rate of 0.008 °/s. Rietveld fitting of Powder X-Ray Diffraction (PXRD) data was completed with the Fullprof Suite software for the initial phase identification.[5–7] The refined lattice parameter for LiYbSe$_2$ is $a$ = 11.26882(4) Å and the refinement parameters $\chi^2$, Rp, R$_{wp}$ and R$_{exp}$ are 2.60, 6.29, 8.50, and 5.28 respectively which indicate a reasonable refinement. The PXRD refinement of LiLuSe$_2$ is given in **Figure S2**.

**Figure S2.** Le-bail refinement of powder diffraction pattern of LiLuSe$_2$. The refined lattice parameter is $a$ = 11.2478 (5) Å in the space group *Fd-3m*.

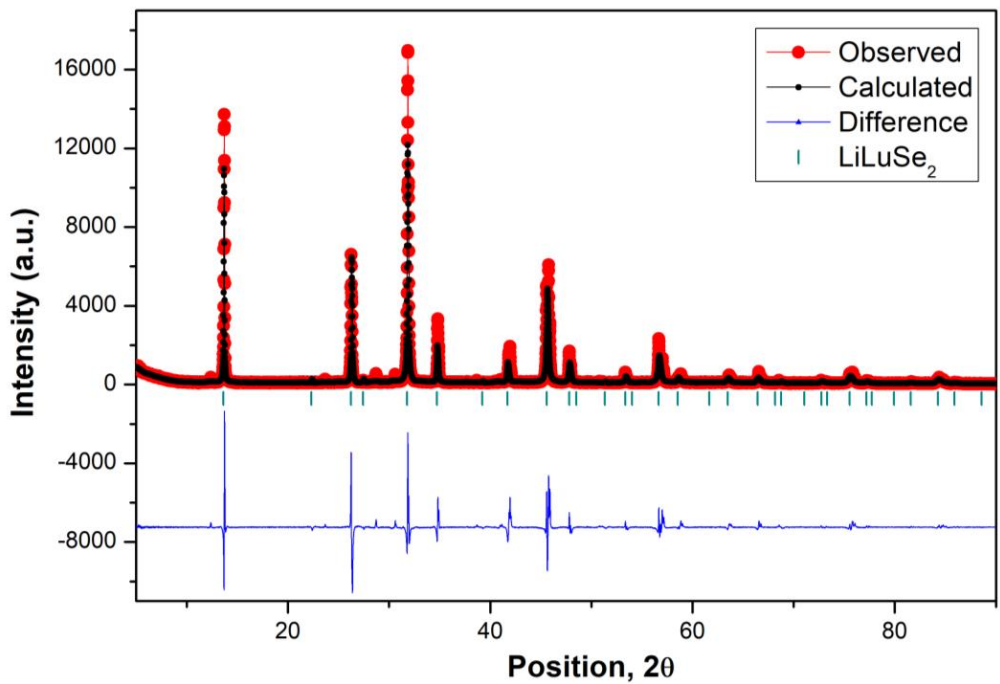

**Magnetic Properties Measured by Physical Property Measurement System (PPMS):** Magnetization measurements of the sample was measured using the Quantum Design Physical Property Measurement (QD-PPMS) Dynacol System. The DC magnetic susceptibility measurements for the polycrystalline LiYbSe$_2$ were conducted in the temperature range from 1.8 K to 300 K under the applied field of 1000 Oe. Magnetic susceptibility is defined as $\chi = M/H$ where M is the magnetization in units of emu/mol and H is the applied field. Magnetic susceptibility data is analyzed by using the modified Curie-Weiss law $\chi = \chi_0 + \frac{C}{T-\theta_{CW}}$, where $\chi$, $\chi_0, C,$ and $\theta_{CW}$ represent magnetic susceptibility, temperature independent contribution to the susceptibility, Curie constant, and the Curie–Weiss temperature, respectively. Detailed results from the Curie-Weiss fitting for both ZFC and FC data are given in the **Figure S3 and Table S4**. Data was fitted for high temperature (HT) and low temperature (LT) regions separately. The field dependence magnetization data was collected at temperatures 2, 10, 20, 150 and 300 K. The heat capacity from 2-100 K was measured on polycrystalline LiYbSe$_2$ and LiLuSe$_2$ using the heat capacity option of QD-PPMS. The heat capacity from 0.07 - 2 K was measured using a dilution refrigerator (DR) of QD-PPMS. A standard relaxation calorimetry method was utilized to conduct the experiment. Our attempt on performing four-probe resistivity measurements in QD-PPMS was unsuccessful as the resistivity of the material was too large to determine from our PPMS. Therefore, resistivity of LiYbSe$_2$ was measured at room temperature using the two-probe method and the resultant resistance was $25\times10^6$ Ω which indicates that LiYbSe$_2$ is an insulator.

**Table S4.** Summary of Curie-Weiss fitting parameters

| μ$_0$H =1000 Oe | Low Temperature (LT) | | | High Temperature (HT) | | |
|---|---|---|---|---|---|---|
| | T range (K) | μ$_{eff}$/μ$_B$ | θ$_{cw}$/ K | T range (K) | μ$_{eff}$/μ$_B$ | θ$_{cw}$/ K |
| ZFC | 2-100 | 3.42 | -8.11 | 200-300 | 4.18 | |
| | 2-10 | 2.94 | -3.14 | | | |
| FC | 2-100 | 3.39 | -8.08 | 200-300 | 3.57 | |
| | 2-10 | 2.93 | -3.11 | | | |

**Figure S3.** Curie-Weiss fitting of inverse magnetic susceptibility (***a***) low temperature (2-10 K) for ZFC (***b***) high temperature (200-300 K) for ZFC. Curie-Weiss fitting of FC data (***c***) 2-10 K (***d***) 200-300 K and (***e***) 2-100 K.

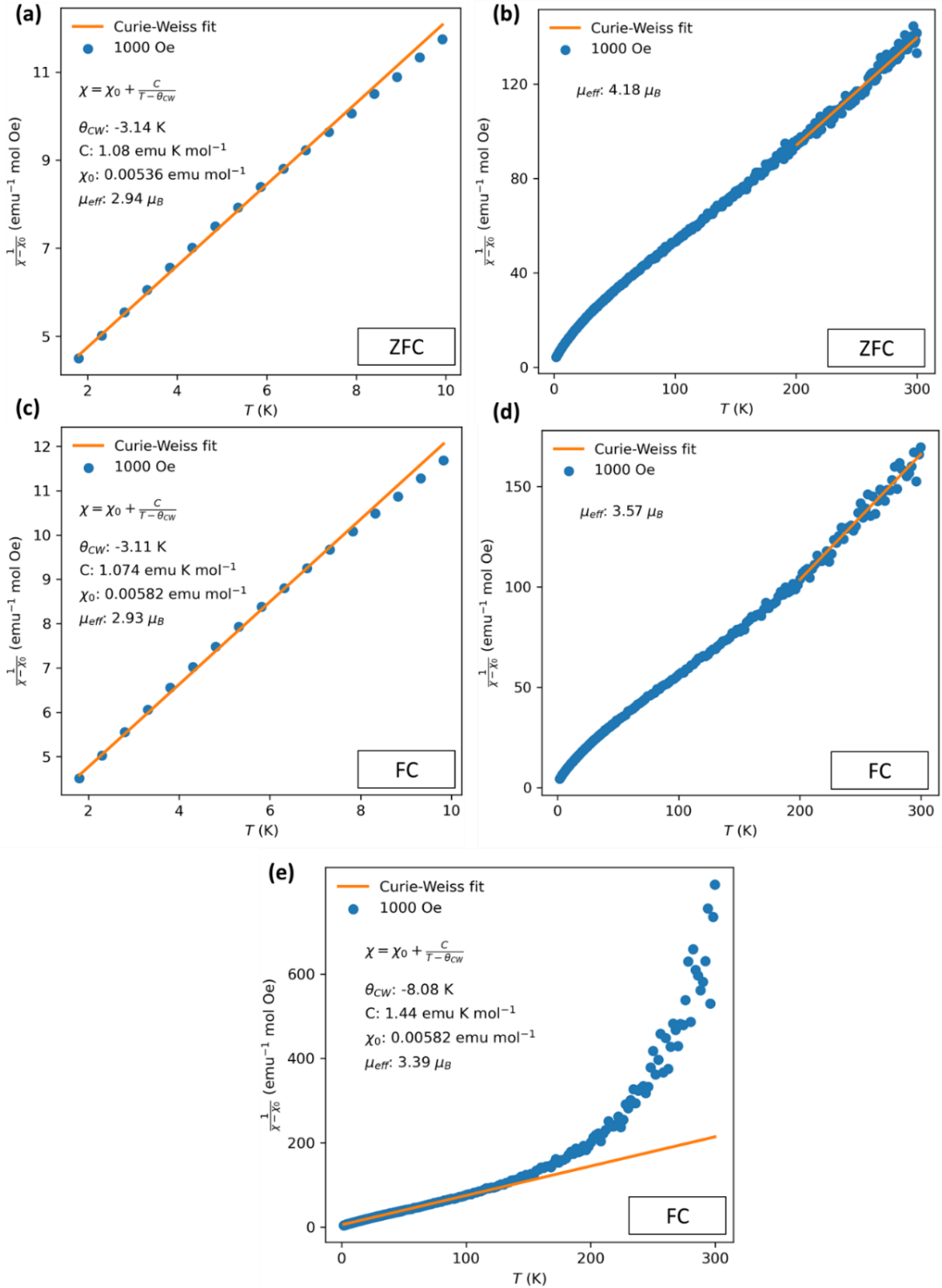

**Table S5:** The fitted parameters with power law for the specific heat of LiYbSe$_2$ under different magnetic fields. For zero field specific heat fitted parameters with both power law and two power law are given.

| | 0 T | 0 T | 0.5 T | 1 T | 4 T | 8 T | 12 T |
|---|---|---|---|---|---|---|---|
| α (J K/mol) | 3.25×10$^{-4}$ | 7.32×10$^{-5}$ | 4.28×10$^{-4}$ | 7.69×10$^{-4}$ | 5.14×10$^{-4}$ | 8.95×10$^{-4}$ | 1.07×10$^{-3}$ |
| b (J/mol K$^{p+1}$) | -0.22 | 0.72 | 1.68 | 1.94 | 0.931 | 0.925 | 0.696 |
| p | 2.29 | 0.52 | 1.25 | 1.77 | 2.03 | 2.45 | 2.48 |
| d (J/mol K$^{q+1}$) | 0.901 | | | | | | |
| q | 0.683 | | | | | | |

**Figure S4.** Temperature dependence of zero field specific heat data fitted with Power law and two power law is compared to describe the magnetic contribution. Within the temperature regime, $T <$ 1 K, no significant differences observed, while two power law indeed fit the data well in the whole 1 K range. All coefficients in the fitting are shown in the figure and summarized in **Table S5**.

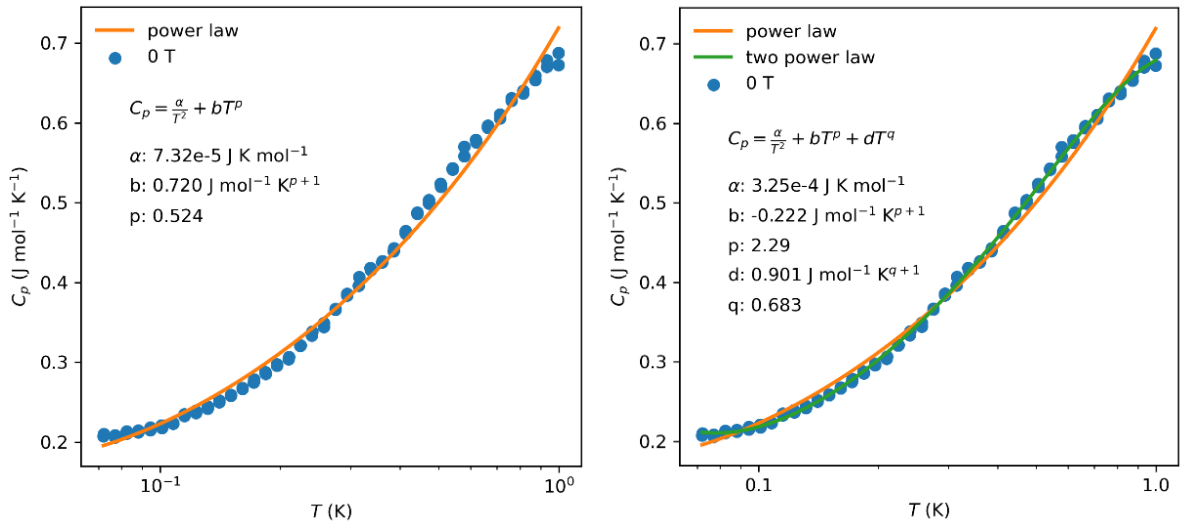

**Figure S5.** Temperature dependence of specific heat data in ultra-low temperature regime under different fields. Data fitted with Power law and the fitted parameters are summarized in the **Table S5**.

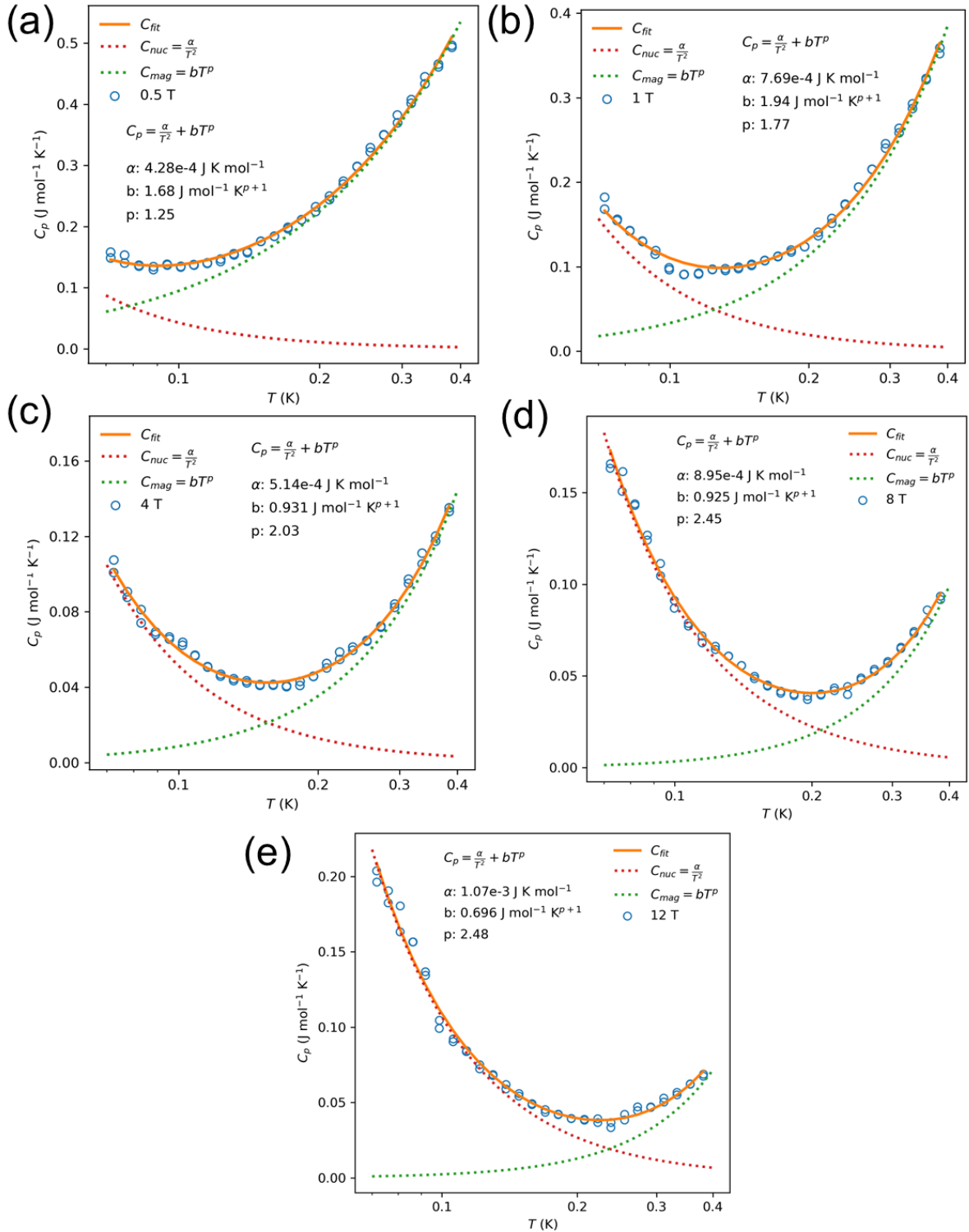

**Figure S6.** Field induced temperature dependence of specific heat at ultra-low temperature regime (***a***) Cp vs T (***b***) C$_{mag}$ vs T (***c***) C$_{mag}$/T vs T and (***d***) cumulative entropy change.

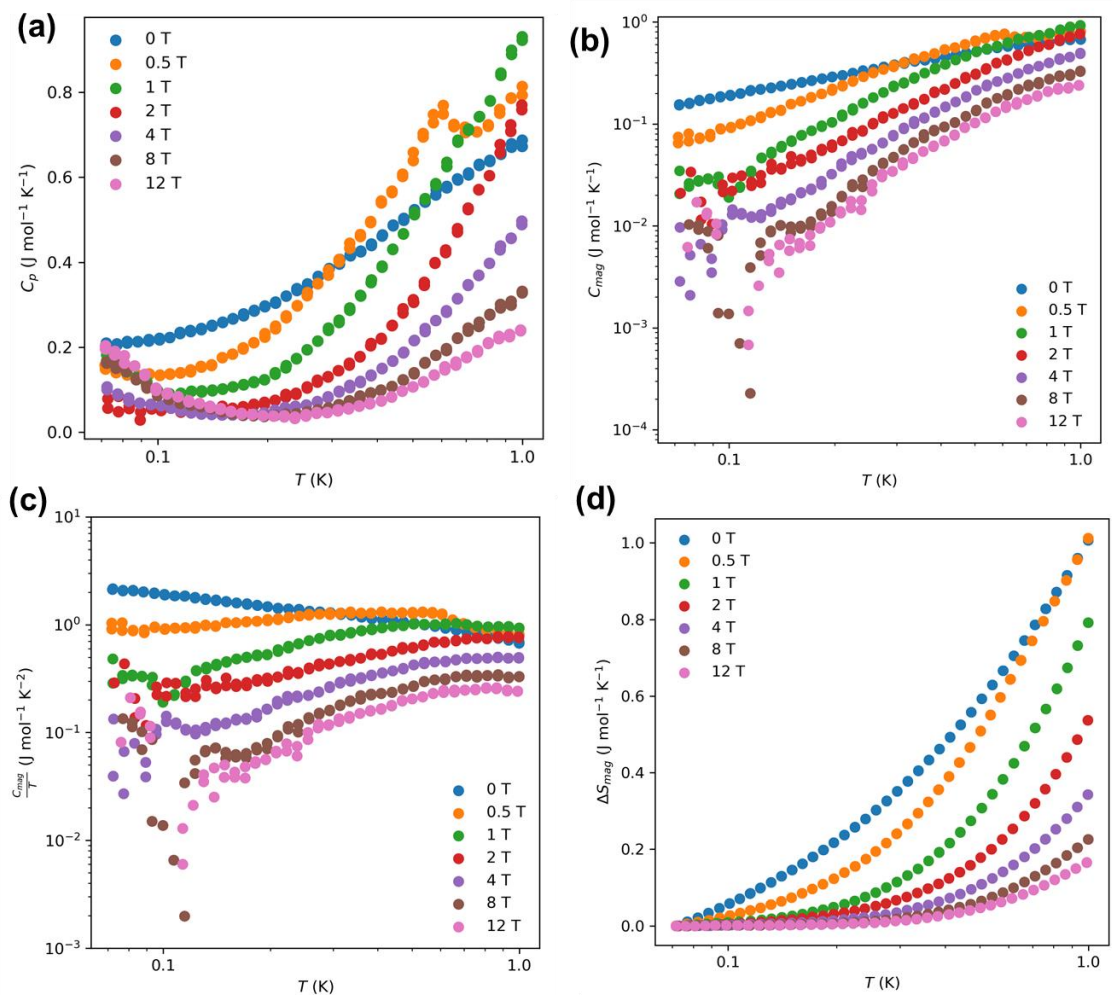

**Figure S7.** (*a*) Field induced temperature dependence of specific heat (*b*) magnetic specific heat and (*c*) magnetic entropy changes over gas constant in the ultra-low + low temperature regime.

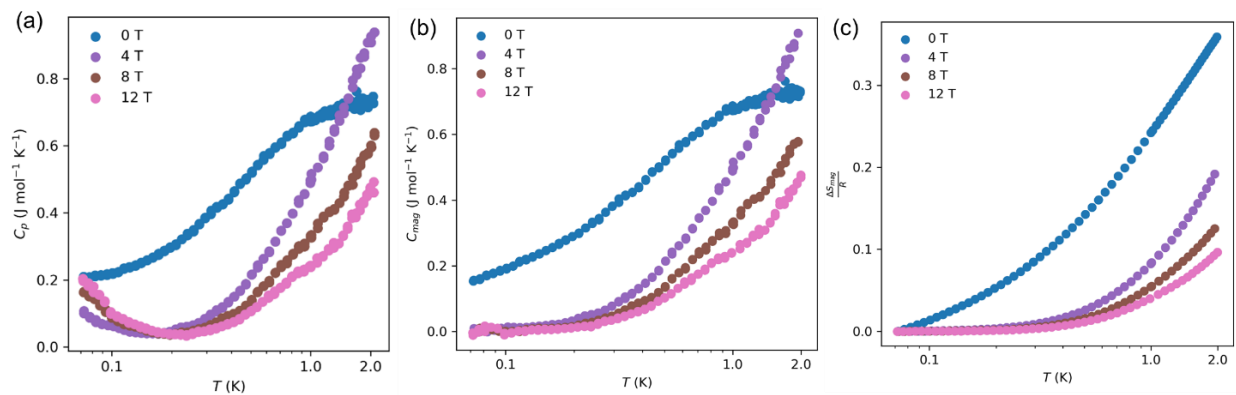

**Figure S8.** Field induced temperature dependence of specific heat in the high temperature regime.

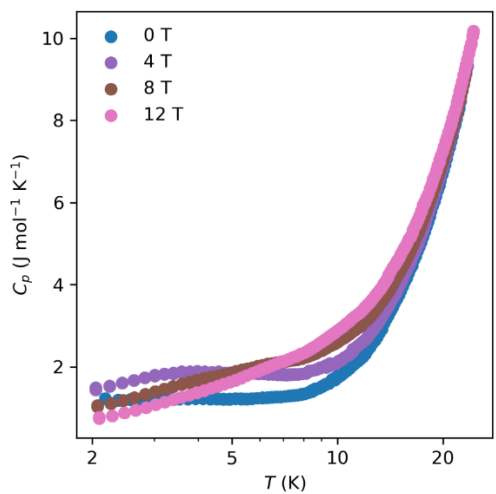

**Electronic Structure Calculation:** The experimental structural parameters obtained from SCXRD were used in the WIEN2k program to calculate the band structure and the density of states (DOS) of the compound LiYbSe$_2$. WIEN2k program operates with the full-potential linearized augmented plane wave method (FP-LAPW) and the implemented local orbitals.[8] The electron correlation within the generalized gradient approximation was handled using the electron exchange-correlation potentials parametrized by Perdew et al.[9] The reciprocal space integrations were completed over a 8 × 8 × 8 Monkhorst-Pack k-points mesh for the non-magnetic calculations.[10] The spin orbit coupling (SOC) effects were applied to both Yb and Se atoms and the spin polarization (SP) from Yb were considered during the calculation. The converged energy per atom was set up to 0.1 meV.

The first-principles calculations on the bulk band structure based on the local-density approximations (LDA) plus correlation parameter (U) and LDA+U with SOC (LDA+U+SOC) methods were performed and results are shown in **Figure S9**. The LDA+U and LDA+U+SOC calculations reveal a calculated semi-metallic ground state contrary to the experimentally realized insulating ground state which needs further theoretical impact to fully understand the system. Based on the LDA calculation, we find that flat and narrow Yb-4*f* bands are located around 0 to -0.5 eV below E$_F$. Contrary to the localized Yb-4*f* states, the valence orbitals from Li and Se atoms seem to strongly hybridize. The Li and Se orbitals contribute two parts – the upper part displays electron-like band dispersion above 2.0 eV, while the lower part lies around −1.0 eV below E$_F$. Moreover, including the SOC effect into the calculation, the Yb-4*f* states were split, resulting in more dispersionless bands distribution through the whole Brillouin zone. Around the Fermi level, the Yb-4*f* bands with a larger band dispersion span and slightly interact with the Se orbitals bands near E$_F$, leading to one flat band at Fermi level. Flat bands can be observed in the geometric frustrated lattices, such as in twisted bilayer graphene and Kagome lattices.[11–13] Moreover, the systems with flat bands can host correlated electronic states, for example, ferromagnetism or superconductivity.[11,14] When the SOC is included in LDA + *U* method with U = 4 eV, the extent of Yb 4*f* electron localization has decreased and a significant split of the band near the fermi level was observed while the Yb and Se states are hybridized and prominent near the fermi level. Due to the conflicting results obtained from our fundamental electronic band structure calculations, it appears that LiYbSe$_2$ in the frustrated pyrochlore structure inherit complex electron behavior.

Therefore, further theoretical studies can help better understand electronic and magnetic properties in terms of pyrochlore structure.

**Figure S9.** Bulk band structure of LiYbSe$_2$ in its non-magnetic phase. (***a***) LDA calculations without the inclusion of spin–orbit coupling (SOC). (***b***) Same as **a,** but with SOC. (***c***) band structure of LiYbSe$_2$ based on LDA + U calculations without SOC. (***d***) Same as **c,** but with SOC.

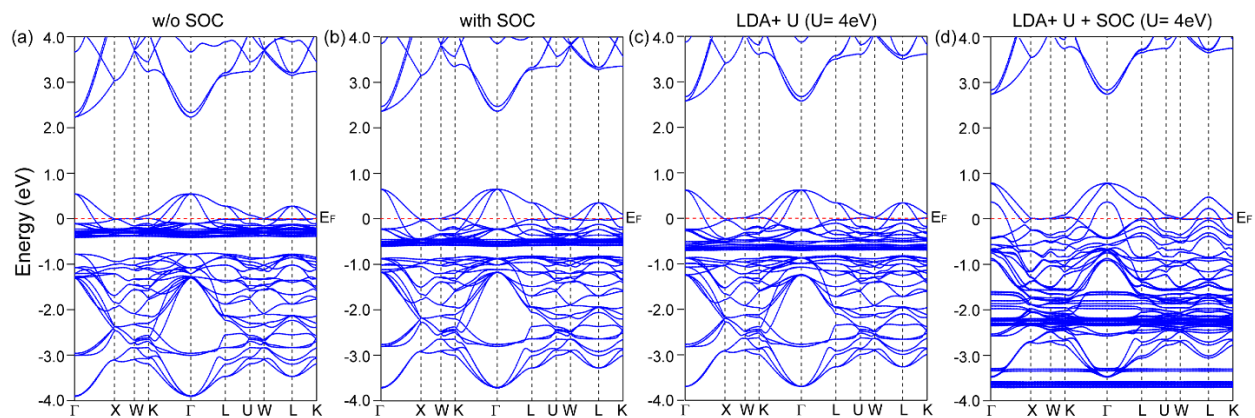